\documentclass[11pt]{article}
\usepackage[a4paper,left=20mm,top=20mm,right=20mm,bottom=25mm]{geometry}

\usepackage{graphicx}
\usepackage{subcaption}
\usepackage{algorithm}
\usepackage[noend]{algpseudocode}
\usepackage{amsmath, amsfonts, amssymb, amsthm}
\usepackage{mathtools}
\usepackage{lscape}

\usepackage{color}
\usepackage[font=footnotesize,width=.85\textwidth,labelfont=bf]{caption}
\usepackage[mathcal]{euscript}
\usepackage{mathptmx} 

\usepackage{authblk}

\newtheorem{thm}{Theorem}[section]%[chapter]

\newtheorem{lem}[thm]{Lemma}
\newtheorem{defn}[thm]{Definition}
\newtheorem*{probl}{Problem}

\renewcommand{\mathbb}{\boldsymbol}

\providecommand{\keywords}[1]{\textbf{\textit{Keywords: }} #1}

\begin{document}


\sloppy

%\title{Characterization of Gene Trees containing HGT and reconcilable Species Trees}
\title{On the Complexity of Recognizing S-composite and S-prime Graphs}

\author[1,2]{Marc Hellmuth}

\affil[1]{Dpt.\ of Mathematics and Computer Science, University of Greifswald, Walther-
  Rathenau-Strasse 47, D-17487 Greifswald, Germany \\
	\texttt{mhellmuth@mailbox.org}}
\affil[2]{Saarland University, Center for Bioinformatics, Building E 2.1, P.O.\ Box 151150, D-66041 Saarbr{\"u}cken, Germany }

\date{}
\normalsize

\maketitle

\abstract{ \noindent
S-prime graphs are graphs that cannot be represented as nontrivial
subgraphs of nontrivial Cartesian products of graphs, i.e., whenever it is
a subgraph of a nontrivial Cartesian product graph it is a subgraph of one
the factors. A graph is S-composite if it is not S-prime. Although linear
time recognition algorithms for determining whether a graph is prime or not
with respect to the Cartesian product are known, it remained unknown if a
similar result holds also for the recognition of S-prime and S-composite
graphs.

In this contribution the computational complexity of recognizing
S-composite and S-prime graphs is considered. Klav{\v{z}}ar \emph{et al.}
[\emph{Discr.\ Math.} \textbf{244}: 223-230 (2002)] proved that a graph is
S-composite if and only if it admits a nontrivial path-$k$-coloring. The
problem of determining whether there exists a path-$k$-coloring for a
given graph is shown to be NP-complete even for $k=2$. This in turn is
utilized to show that determining whether a graph is S-composite is
NP-complete and thus, determining whether a graph is S-prime is
CoNP-complete. Many other problems are shown to be NP-hard, using
the latter results. 
}

\bigskip
\noindent

\sloppy

\maketitle

%\end{keyword}

\section{Introduction and Preliminaries}

A graph $S$ is said to be \emph{S-prime} (\emph{S} stands for ``subgraph'')
w.r.t. to an arbitrary graph product $\star$ if for all graphs $G$ and $H$
with $S\subseteq G\star H$ holds: $S\subseteq H$ or $S\subseteq G$. A graph
is \emph{S-composite} if it is not S-prime. The only S-prime graphs w.r.t.\
the direct product are complete graphs or complete graphs minus an edge
\cite{Sab:75}. The only S-prime graphs w.r.t.\ the strong product and the
lexicographic product are the single vertex graph $K_1$, the disjoint union
$K_1\cup K_1$ and the complete graph on two vertices $K_2$ \cite{LB:81,
LB:95}.

Not much is known, however, about the structure of S-prime graphs w.r.t.\
the Cartesian product. Examples include complete graphs $K_n$ with
$n\geq 1$ vertices and complete bipartite graphs $K_{m,n}$ with $m\geq
2, n\geq 3$. Another class of Cartesian S-prime graphs are so-called
diagonalized Cartesian products of S-prime graphs \cite{HOS12}, which in
turn play an important role in finding approximate strong product graphs,
see \cite{Hellmuth:11} 
or to find the prime factors of so-called hypergraphs \cite{HON-14,HL:16}.
Several interesting characterizations of (basic)
S-prime graphs due to Lamprey and Barnes \cite{LB:81, LB:95}, Klav{\v{z}}ar
\emph{et al.} \cite{Klavzar:02, KP:05} and Bre{\v{s}}ar \cite{Bresar:03}
are known. However, although those characterizations are established and
graphs can be recognized as prime (or factorizable) w.r.t. the Cartesian
product in linear time \cite{IP:07}, it remained unknown if a similar
result holds also for the recognition of S-prime and S-composite graphs. We
will show in this contribution that the problem of determining whether a
graph is S-composite, resp., S-prime w.r.t.\ the Cartesian product is
NP-complete, resp., CoNP-complete. Moreover, using this result we are able
to show the NP-hardness of several other problems. 

Before we proceed, we introduce some notation. We consider finite, simple,
connected and undirected graphs $G=(V,E)$ with vertex set $V$ and edge set
$E$. A graph $H$ is a \emph{subgraph} of a graph $G$, in symbols
$H\subseteq G$, if $V(H)\subseteq V(G)$ and $E(H)\subseteq E(G)$. If
$H\subseteq G$ and all pairs of adjacent vertices in $G$ are also adjacent
in $H$ then $H$ is called an \emph{induced} subgraph. The subgraph of a
graph $G$ that is induced by a vertex set $W \subseteq V(G)$ is denoted by
$\langle W \rangle$. We define the \emph{neighborhood} of vertex $v$ as the
set $N(v) = \{x\in V(G)\mid (v,x) \in E\}$. A \emph{complete graph} $K_n$
is a graph in which every vertex is adjacent to every other vertex. 

In the following we will consider the Cartesian product only. Therefore,
the terms S-prime and S-composite refer to this product from here on. 

The Cartesian product $G\Box H$ has vertex set $V(G\Box H)=V(G)\times
V(H)$; two vertices $(g_1,h_1)$, $(g_2,h_2)$ are adjacent in $G\Box H$ if
$(g_1,g_2)\in E(G)$ and $h_1=h_2$, or $(h_1,h_2)\in E(G_2)$ and $g_1 =
g_2$. The one-vertex complete graph $K_1$ serves as a unit, as $K_1 \Box H
= H$ for all graphs $H$. A Cartesian product $G\Box H$ is called
\emph{trivial} if $G \simeq K_1$ or $H \simeq K_1$. A graph $G$ is
\emph{prime} with respect to the Cartesian product if it has only a trivial
Cartesian product representation. The Cartesian product is associative.
Therefore, a vertex $x$ of a Cartesian product $\Box_{i=1}^n G_i$ is
properly ``coordinatized'' by the vector $c(x) := (c_1(x),\dots,c_n(x))$
whose entries are the vertices $c_i(x)$ of its factor graphs $G_i$. Two
adjacent vertices in a Cartesian product graph therefore differ in exactly
one coordinate. The Cartesian product $Q_n = \Box_{i=1}^n K_2$ is called
\emph{hypercube}. W.l.o.g. we assume that $c(x)\in \{0,1\}^n$ for all $x
\in V(Q_n)$. For detailed information about product graphs we refer the
interested reader to \cite{Hammack:2011a,IMKL-00} or \cite{IMKLDO-08}.

\begin{figure}[htbp]
  \centering
  \includegraphics[bb= 95 600 401 665]{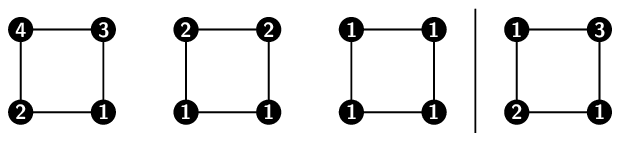}
  \caption{Possible path-$k$-coloring of a square $Q_2$ for $k=1,2,4$.  A
    possible well-coloring that is not a path-3-coloring is shown on the
    right-hand side graph \cite{HOS12}. }
  \label{fig:k-color-Q2}
\end{figure}

For our purposes, the characterization of S-composite graphs in terms of
particular colorings \cite{Klavzar:02} is of direct interest. A
\emph{$k$-coloring} of $G$ is a surjective mapping
$\mathcal{C}:V(G)\rightarrow\{1,\dots,k\}$. This coloring need not be
proper, i.e., adjacent vertices may obtain the same color. A path $P$ in
$G$ is \emph{well-colored} by $\mathcal{C}$ if for any two consecutive
vertices $u$ and $v$ of $P$ we have $\mathcal{C}(u) \neq \mathcal{C}(v)$.
Following \cite{Klavzar:02}, we say that $\mathcal{C}$ is a
\emph{path-$k$-coloring} of $G$ if $\mathcal{C}(u)\neq \mathcal{C}(v)$
holds for the endpoints of every well-colored $u,v$-path $P$ in $G$, see
Figure~\ref{fig:k-color-Q2}. For $k=1$ and $k=|V|$ there are trivial
path-$k$-colorings: For $k=1$ the coloring is constant and hence there are
no well-colored paths. On the other hand if a different color is used for
every vertex, then every path, of course, has distinctly colored
endpoints. A path-$k$-coloring is nontrivial if $2\leq k \leq |V(G)|-1$.

\begin{thm}[\cite{Klavzar:02}]
  A connected graph $G$ is S-composite if and only if there exists a nontrivial 
  path-$k$-coloring of $G$.
  \label{thm:S-composite}
\end{thm}

\section{Results}

We start our exposition with some known results on path-$k$-colorings. We
continue to examine basic properties of path-$2$-colorings of hypercubes
and so-called joint graphs as well as properties of path-$k$-colorings of
so-called $k$-extended joint graphs. These results will be used to show
that the problem of determining whether there exists a path-$k$-coloring
for a given graph is NP-complete. Hence, there is no polynomial time
algorithm solving this problem unless $P=NP$. Using the latter result we
can easily infer the NP-completeness of the problem to decide whether a
graph is S-composite and therefore, we can conclude that determining
whether a graph is S-prime is CoNP-complete. Moreover, other decision
problems concerning particular graph properties that are equivalent to
decide whether a graph is S-composite are therefore NP-hard. Those problems
are stated at the end of this section. For detailed information about
(Co)NP-completeness we refer the interested reader to \cite{GJ79}.

\subsection{Path-$k$-colorings, Hypercubes, the Joint Graph $G(m)$
						 and the $k$-Extended Joint Graph $G(m,k)$}

\begin{lem}[\cite{HOS12}]
  Let $H\subseteq G$ and suppose $\mathcal{C}$ is a path-$k$-coloring of $G$.
  Then the restriction $\mathcal{C}_{|V(H)}$ of $\mathcal{C}$ on $V(H)$ 
  is a path-$l$-coloring of $H$, $l\leq k$.
  Moreover, if $V(H)=V(G)$ and $\mathcal{C}$ is a nontrivial path-$k$-coloring
  of $G$, then it is also  a nontrivial path-$k$-coloring of $H$.
  \label{lem:subgraph-k-coloring}
	 \label{subgraohW.col}
\end{lem}

\begin{lem}[\cite{HOS12}]
  Let $\mathcal{C}$ be a path-$k$-coloring of the Cartesian product 
  $G = \Box_{i=1}^n S_i$ of S-prime graphs $S_i$ 
  and suppose there are two vertices 
  with maximal distance in $G$ that have the same
  color. Then $\mathcal{C}$ is constant on $G$, i.e., $k=1$. 
  \label{cor:max-dist-uv-sameColor}
\end{lem}

\begin{lem}[\cite{HOS12}]
  The hypercube $Q_2 = K_2 \Box K_2$ has no path-$3$-coloring. In particular, every 
  path-$2$-coloring of $Q_2$ has adjacent vertices with the same color. 
  \label{lem:G-k-color}
\end{lem}

From Theorem $9$ 
(Path-$k$-coloring of Cartesian products of S-prime Graphs)
in \cite{HOS12} we can easily derive the next lemma.
\begin{lem}
	\label{lem:p24kQ3}
	 Any nontrivial path-$k$-coloring of a hypercube $Q_3$ 
	is either a path-$2$-coloring or a path-$4$-coloring.
\end{lem}

\begin{figure}[tbp]
  \centering
  \includegraphics[bb= 261 555 439 710]{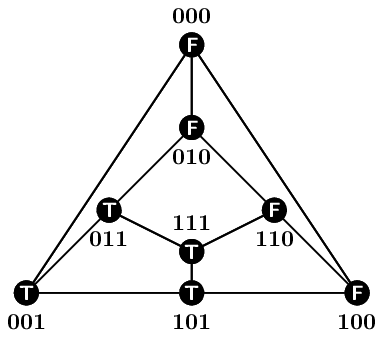}
  \caption{Shown is a hypercube $Q_3$, together with its coordinatized vertices
  				and a possible path-$2$-coloring.}
  \label{fig:k-color-Q3}
\end{figure}

In the following lemma a simple but useful result concerning
nontrivial path-$k$-colorings of hypercubes $Q_3$ is proved. 
%It is shown that for any
%path-$2$-coloring $\mathcal{C}$ of $Q_3$ any vertex $x\in Q_3$ has exactly
%one neighbor $w\in N(x)$ with $C(w)\neq C(x)$ and that two diametrical
%vertices have different colors. 

\begin{lem}
	\label{lem:p2kQ3-1}
	\label{lem-diametricalVdiffC}
	  Let $\mathcal{C}$ be a nontrivial path-$k$-coloring of the hypercube $Q_3 =
	  (V,E)$. Then $\mathcal{C}(x)\neq \mathcal{C}(y)$ for all vertices $x,y
	  \in V$ with distance $d(x,y)=3$.

		If $k=2$ then for all vertices $x\in V$ there
	  are two different vertices $u,v \in N(x)$ with color
	  $\mathcal{C}(u)=\mathcal{C}(v)=\mathcal{C}(x)$ and one vertex $w\in
	  N(x)$ with $\mathcal{C}(w) \neq \mathcal{C}(x)$.

		If $k=4$ then for all vertices $x\in V$ there is 
		one vertex $u\in N(x)$ with color $\mathcal{C}(u)=\mathcal{C}(x)$ and
		there is a well-colored path from $u$ to $y$, where $y\in V$ with $d(x,y)=3$.
\end{lem}

\begin{proof}
Let $x$ and $y$ be two vertices with maximal distance $d(x,y)=3$. Since
$Q_3$ is the Cartesian product of S-prime graphs and $\mathcal{C}$ is a
nontrivial path-$k$-coloring, Lemma \ref{cor:max-dist-uv-sameColor} implies that
$\mathcal{C}(x) \neq \mathcal{C}(y)$.

Let $\mathcal{C}$ be a path-$2$-coloring of the hypercube $Q_3$. 
Let $N(x)=\{u,v,w\}$ be the set of adjacent vertices to $x$ and $y$ be the
vertex with maximal distance $d(x,y)=3$ to $x$. Due to the three paths
$\langle\{v,x,w\}\rangle$, $\langle\{v,x,u\}\rangle$ and
$\langle\{u,x,w\}\rangle$, we can conclude that there is at most one vertex
in $N(x)$ with color $\mathcal{C}(y)$, otherwise $\mathcal{C}$ would not be
a path-$2$-coloring. Assume that all vertices $u,v,w$ have color
$\mathcal{C}(x)$. Consider the squares $\langle\{x,u,a,v\}\rangle$ and
$\langle\{x,u,b,w\}\rangle$ in $Q_3$. Lemma \ref{lem:subgraph-k-coloring}
and \ref{lem:G-k-color} imply that $\mathcal{C}(a)
=\mathcal{C}(b)=\mathcal{C}(x)$. Since $\mathcal{C}(x)\neq \mathcal{C}(y)$
we can conclude that the path $a-y-b$ is well-colored, contradicting that
$\mathcal{C}$ is a path-$2$-coloring. Thus, exactly one vertex contained in
$N(x)$ has color $\mathcal{C}(y)$ and the other two have color
$\mathcal{C}(x)$.

Let $\mathcal{C}$ be a path-$4$-coloring of the hypercube $Q_3$. Assume
there is a vertex $x\in V$ such that for each vertex $z \in N(x)=\{u,v,w\}$
holds $\mathcal C(z) \neq C(x)$. Note, in this case it holds $\mathcal C(z)
\neq \mathcal C(z')$ for each $z, z'\in N(x)$ with $z\neq z'$, otherwise
there is a well-colored path $z-x-z'$, a contradiction. Consider the square
$\langle\{x,u,a,v\}\rangle$ in $Q_3$. Thus, by Lemma
\ref{lem:subgraph-k-coloring} and Lemma \ref{lem:G-k-color} the vertex $a$
must obtain the fourth color, namely $\mathcal C(w)$. Hence, there is a
well-colored path $a-u-x-w$ with $\mathcal C(w) = \mathcal C(a)$, a
contradiction. Therefore, there is a vertex $u\in N(x)$ with color
$\mathcal{C}(u)=\mathcal{C}(x)$. It remains to show that there is a
well-colored path from $u$ to $y$, where $y\in V$ with $d(x,y)=3$. Note,
we have $\mathcal{C}(u)\neq\mathcal{C}(y)$. Assume there is no well-colored
path from $u$ to $y$. Thus, the two shortest paths $u-z-y$ and $u-z'-y$ are
not well-colored and therefore, $\mathcal{C}(z), \mathcal{C}(z') \in \{
\mathcal{C}(u), \mathcal{C}(y)\}$. If the vertices $z$ and $z'$ have both
color $\mathcal{C}(u)$, resp., $\mathcal{C}(y)$ there is a well-colored
path $z-y-z'$, resp., $z-u-z'$, a contradiction. Thus, one vertex must have
color $\mathcal{C}(u)$, say the vertex $z$, while the other vertex $z'$ has
color $\mathcal{C}(y)$. Consider the square $\langle\{x,u,z,v\}\rangle$ and
the square $\langle\{x,u,z',v'\}\rangle$ in $Q_3$. By construction and
Lemma \ref{lem:G-k-color}, vertex $v$ must get color $\mathcal C(x)$ and
vertex $v'$ the color $\mathcal C(y)$. Again, Lemma \ref{lem:G-k-color}
implies that the vertex $a$ contained in the square
$\langle\{v,z,y,a\}\rangle$ must get color $\mathcal C(y)$ and we obtain a
valid path-$2$-coloring, a contradiction since we assumed to have a
path-$4$-coloring. Therefore, in any path-$4$-coloring one of the vertices
$z$ or $z'$ gets a color different from $\mathcal{C}(u)$ and
$\mathcal{C}(y)$ and hence, one of the paths $u-z-y$ or $u-z'-y$ is
well-colored.
\end{proof}

\begin{figure}[tbp]
  \centering
  \includegraphics[bb= 37 456 572 758, scale=0.8]{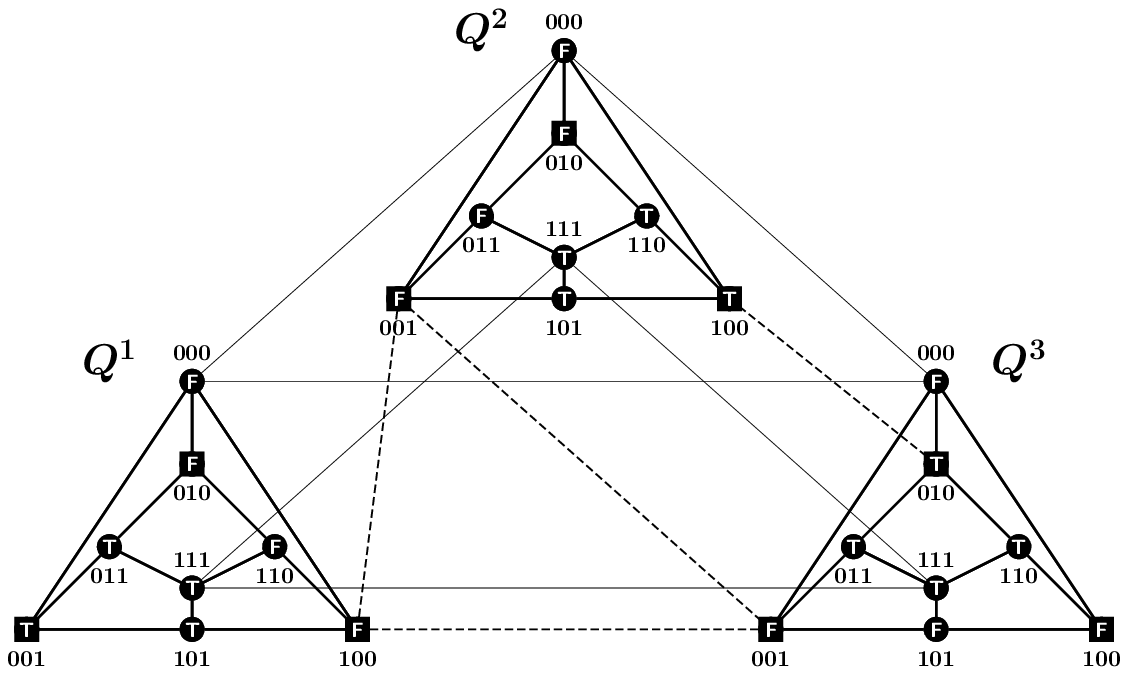}
  \caption{Shown is a joint graph $G(m)$ based on the set
           $\mathcal{Q}=\{Q^1,Q^2,Q^3\}$ together with a path-$2$-coloring
           $\mathcal{C}: V(G(m)) \rightarrow \{\mathbb{F},\mathbb{T}\}$. 
  				In particular, $G(m)$ reflects the monotone 1-in-3 SAT formula 
  				$\psi = \{L_1, L_2, L_3\}$ with clauses $L_1=(b_1,b_2,b_3), L_2=(b_1,b_4,b_5)$ and 
  				$L_3=(b_1,b_4,b_6)$. Each clause $L_i$ is identified with the hypercube $Q^i$.
				Each of the variables $b_{i_1},b_{i_2},b_{i_3}\in L_i$ is uniquely identified 
				with one of the vertices in $V(Q^i)$ that have coordinates $(001)$, $(010)$, resp., $(100)$ 
				(highlighted by square vertices). 
				Edges $(u,v)$ between different hypercubes are added, whenever
				$c^j(u)=c^i(v)\in \{(000),(111)\}$ (thin-lined edges) 
				or the Boolean variables the 
				vertices are associated to are identical (dashed-lined edges).}   		
  \label{fig:jointgraph}
\end{figure}

\begin{figure}[tbp]
  \centering
  \includegraphics[bb= 23 344 526 790, scale=0.6]{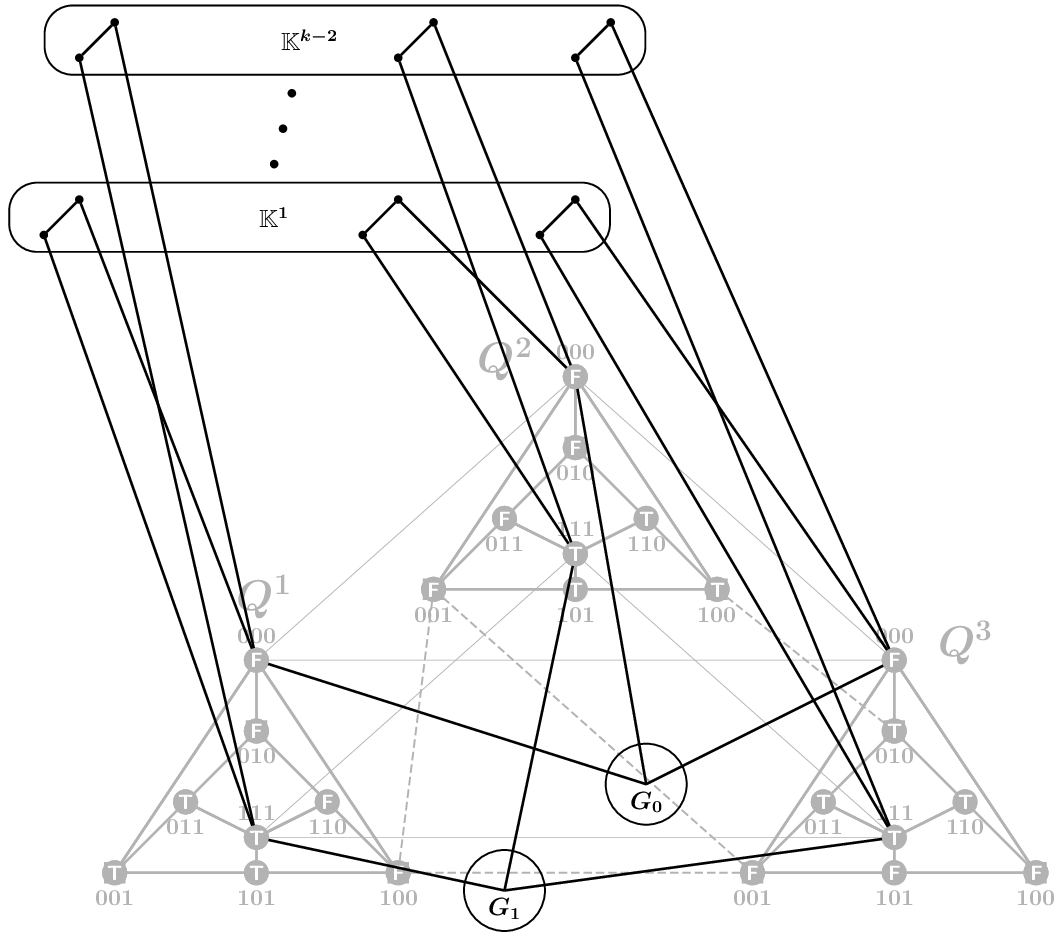}
  \caption{Shown is the $k$-extended joint graph $G(m, k)$ for the joint graph
						$G(m)$ shown in Fig. \ref{fig:jointgraph}. The graphs $G_0$ and 
						$G_1$ are complete graphs of size $k-m+1$. Each vertex of $G_i$
						is connected to each vertex with respective coordinates $(iii)$ for
						$i=0,1$. Thus, the vertices with respective coordinates $(iii)$ 
						together with the vertices of $G_i$ induce a complete subgraph 
						of size $k+1$. 
						Finally, $k-2$ disjoint complete graphs $\mathbb K^i\in \mathcal K$
						of size $2m+k+1$ are added to $G(m)$. 
						Vertices $v$ and $w$ with coordinates $(000)$, resp., $(111)$ 
						contained in each $Q^j\in \mathcal Q$ 
						are connected to arbitrary but different vertices 
						$x, y \in V(\mathbb K^i)$ for each $\mathbb K^i\in \mathcal K$. 
						Those edges $(v,x)$ and $(w,y)$ are summarized in the set $F'$. } 
  \label{fig:extjointgraph}
\end{figure}

Now, we establish the so-called joint graph $G(m)$ and the $k$-extended 
joint graph $G(m,k)$, that will become a powerful tool
for proving the NP-hardness, as we shall see later; 
 see also also Figure \ref{fig:jointgraph} and \ref{fig:extjointgraph}.

\begin{defn}[Joint Graph $G(m)$]%[Joint Graph $\boldsymbol{G(m)}$]
Let $\mathcal{Q}=\{Q^1,\dots,Q^m\}$ be a set of hypercubes $Q_3$ and let
the respective coordinates of vertices $v\in V(Q^j)$ be denoted by
$c^j(v)\in\{0,1\}^3$. Based on the set $\mathcal{Q}$ we define the
\emph{joint graph $G(m) = (V,E)$} as follows:

$$V=\cup_{i=1}^m V(Q^i)$$ and 
$$E=\cup_{i=1}^m E(Q_i)\cup E_{0} \cup E_{1} \cup E',$$
where for $k\in \{0,1\}$ the edge set $E_k$ is defined as the set $\{(u,v)
\mid c^j(u)=c^i(v)=(kkk), i\neq j\}$ and $E'$ denotes a set of arbitrarily
added edges between vertices $u\in V(Q^j)$ and $v \in V(Q^i)$ with $i\neq j$
and 
$c^j(u),c^i(v)\in \{(001),(010),(100)\}$  with the restriction
that each such vertex $u \in V(Q^i)$ is connected to at most one other
vertex $v\in V(Q^j)$ for each $j$. 
\label{def:Gm}
\end{defn}
 
In other words, the joint graph $G(m)$ consists of $m$ disjoint copies of
hypercubes $Q_3$, where all pairwise different vertices with coordinates
$(000)$ are forced to be adjacent resulting in a complete subgraph $K_m$. These edges
are contained in $E_0$. In the same way vertices with coordinates $(111)$
are connected by edges contained in $E_1$ and thus, the induced subgraph of
these vertices results in a complete subgraph $K_m$ as well. Moreover, additional
edges contained in $E'$ between vertices with coordinates $(001)$, $(010)$
and $(100)$ between different copies of these hypercubes can be placed in a
restricted way; see also Figure \ref{fig:jointgraph}.

\begin{defn}[$k$-Extended Joint Graph $G(m,k)$]%[Joint Graph $\boldsymbol{G(m)}$]
Let $G(m)$ be a given joint graph, $k$ be an arbitrary integer and
$V_0, V_1\subset V(G(m))$ be the set of vertices with coordinates $(000)$ 
and $(111)$ in $G(m)$, respectively. Moreover, let 
$\mathcal K = \left\{ \mathbb K^1, \dots, \mathbb  K^{k-2}\right\}$
be the set of $k-2$ complete subgraphs of size $2m+k+1$.
% and $E_m(K^l)=\{e_1,\dots e_m\}\subseteq E(K^l)$ be a set
%of $m$ disjoint non-incident edges of $K^l$, $l=1,\dots,k-2$.
With $G_0 =(W_0,E(G_0))$ and $G_1=(W_1,E(G_1))$ we denote two further
disjoint complete graphs of size $k-m+1$, while $G_0 = G_1 = \emptyset$
if $k-m+1\leq 0$.
The \emph{$k$-extended joint graph} $G(m,k) = (V,E) $ consists of
$$V= \cup_{i=1}^{k-2} V(\mathbb K^{i}) \cup V(G(m)) \cup W_0 \cup W_1$$ and 
$$E= \cup_{i=1}^{k-2} E(\mathbb K^{i}) \cup E(G(m)) \cup E(G_0) \cup E(G_1) \cup F_{0} \cup F_{1} \cup F'$$ 
where for $i\in \{0,1\}$ the edge set $F_i$ is defined as the set $\{(u,v)
\mid u\in V_i, v\in W_i \}$.  
The edges in $F'$ connect all vertices $v\in V_0\cup V_1$
%V(G(m))$ with coordinates  $(000)$ or $(111)$ 
to vertices  $w\in V(\mathbb K^i)$, while different vertices
of $G(m)$ are connected to different vertices  $w\in V(\mathbb K^i)$, 
for each $i=1,\dots, k-2$. 
\label{def:extGm}
\end{defn}

In other words, the $k$-extended joint graph $G(m,k)$ consists of the graph
$G(m)$, $k-2$ disjoint copies of complete graphs $\mathbb K^j\in \mathcal
K$ of size $2m+k+1$ and two disjoint copies of complete graphs $G_0=$ and
$G_1$ of size of size $k-m+1$. All pairwise different vertices with
coordinates $(000)$ and $(111)$ are connected to pairwise different
vertices in $V(\mathbb K^j)$ for each $\mathbb K^j\in \mathcal K$. These
edges are contained in $F'$. Hence, any pair of vertices $(u,v)\in V(Q^i)$
with coordinates $(000)$ and $(111)$ is mapped to an edge in $\mathbb K^j$,
while such pairs of vertices of different hypercubes are mapped to
different non-incident edges in $\mathbb K^j$ for each $\mathbb K^j\in
\mathcal K$. The edges in $F_i$ connect all vertices with coordinates
$(iii)$ to all vertices contained in $G_i$, $i=0,1$. Hence, $\langle V_0
\cup W_0 \rangle \simeq \langle V_1 \cup W_1 \rangle \simeq K_{k+1}$. Note,
$G(m) = G(m,2)$ for $m>2$.

\begin{lem}
	\label{lem:p2kG=UnionQ3}
	Let $G(m)$ be a joint graph for the set of hypercubes $\mathcal{Q}$ and let $m>2$.
	If $\mathcal{C}$ is a path-$2$-coloring of $G(m)$ then all vertices $v$
	and $w$ connecting different hypercubes $Q^i, Q^j \in \mathcal Q$, i.e.,
	$(v,w)\in E_0 \cup E_1 \cup E'$, must have the same color
	$\mathcal{C}(v)=\mathcal{C}(w)$.
	Furthermore, $\mathcal{C}(v)\neq\mathcal{C}(w)$ for all vertices $v,w$
	with $c^j(v) = (000)$ and $c^i(w) = (111)$, $i,j\in\{1,\dots,m\}$.
\end{lem}

\begin{proof}
	First, we show that all vertices with $c^j(v) = (000)$, resp., $c^j(w) =
	(111)$ share the same color, while $\mathcal{C}(v)\neq\mathcal{C}(w)$
	for all $v,w$ with $c^j(v) = (000)$ and $c^i(w) = (111)$. Since all
	vertices $v,v'$ with coordinates $(000)$ are connected via edges $(v,v')
	\in E_0$ they induce a complete subgraph $K_m$. Since $m>2$, we can
	conclude that $\mathcal{C}(v) = \mathcal{C}(v')$, otherwise
	$\mathcal{C}$ would not be a path-$2$-coloring of $G(m)$. 
	Analogously, we have $\mathcal{C}(w') = \mathcal{C}(w)$ for all vertices
	$w,w'$ with coordinates $(111)$. 
	
	Now, assume for contradiction that $\mathcal{C}(v) = \mathcal{C}(w)$ for
	some vertex $v$ with coordinates $c^j(v) = (000)$ and $w$ with $c^i(w) =
	(111)$. Using the previous arguments we can conclude that for all
	$Q^j\in \mathcal{Q}$ it is the case that $\mathcal{C}(v') =
	\mathcal{C}(w')$ for $v', w' \in V(Q^j)$ with $c^j(v') = (000)$ and
	$c^j(w') = (111)$. Lemma \ref{lem:subgraph-k-coloring} and
	\ref{lem:p2kQ3-1} imply that all vertices within the induced subgraph
	$\langle V(Q^j)\rangle$ must have same color, this holds for all $Q^j\in
	\mathcal{Q}$ and therefore $\mathcal{C}(v) = \mathcal{C}(w)$ for all
	$v,w \in V(G(m))$, contradicting that $\mathcal{C}$ is a
	path-$2$-coloring of $G(m)$.
   
   It remains to show that for vertices $a,b$ with $(a,b)\in E'$ it is the
   case that $\mathcal{C}(a)=\mathcal{C}(b)$. Let $a\in V(Q^j)$ and $b\in
   V(Q^i)$. Since $\mathcal{C}(v) \neq \mathcal{C}(w)$ for $v, w \in
   V(Q^j)$ with $c^j(v) = (000)$ and $c^j(w) = (111)$, it follows from
   Lemma \ref{lem:subgraph-k-coloring} that each induced subgraph $\langle
   V(Q^j) \rangle$ must be ``proper'' path-$2$-colored. Thus, Lemma
   \ref{lem:p2kQ3-1} implies that there is a vertex $a'\in N(a)\cap V(Q^j)$
   with $\mathcal{C}(a') \neq \mathcal{C}(a)$. Hence, if $\mathcal{C}(b)
   \neq \mathcal{C}(a)$ the path $a'-a-b$ would be
   well-colored but $\mathcal{C}(a') = \mathcal{C}(b)$, a contradiction. 
 \end{proof}

\begin{lem}
	Let $\mathcal C$ be a nontrivial path-$k$-coloring of the $k$-extended joint graph $G(m,k)$, $k>2$. 
	Then $\mathcal C_{|V(G(m))}$ is a path-$2$-coloring of the underlying joint graph $G(m)$.
	\label{lem:kcol-to-2col}
\end{lem}
\begin{proof}
	Let $\mathcal C$ be a nontrivial path-$k$-coloring of $G(m,k)$, $k>2$.
	First we show that all vertices $v\in V(Q^i)$ and $w\in V(Q^j)$ with
	coordinates $c^i(v) = (000)$ and $c^j(w) = (111)$ must have different
	colors, $\mathcal C(v)\neq \mathcal C(w)$. Assume $\mathcal C(v) =
	\mathcal C(w)$ for some vertices $v \in V(Q^i)$ with coordinates $(000)$
	and $w\in V(Q^j)$ with coordinates $(111)$. Since the vertices in $G_0$
	are connected to all vertices with coordinates $(000)$ and since those
	vertices together with $G_0$ induce a complete graph $K_{k+1}$ all
	vertices within $G_0$ and all vertices with coordinates $(000)$ must have
	the same color. Analogously, all vertices within $G_1$ and all vertices
	with coordinates $(111)$ must have the same color. Therefore, $\mathcal
	C(v) = \mathcal C(w)$ for all vertices with coordinates $(000)$ and
	$(111)$. Lemma \ref{subgraohW.col} and \ref{lem-diametricalVdiffC} imply
	that each $Q^i \in \mathcal Q$ can then only be trivial path-$k$-colored,
	i.e., either path-$1$-colored or path-$8$-colored. Since we assumed $C(v)
	= \mathcal C(w)$ each $Q^i \in \mathcal Q$ must be path-$1$-colored and
	hence, all vertices within each $Q^i \in \mathcal Q$ have the same
	color. Therefore, the subgraph $\langle \cup_{i=1}^m V(Q^m)\cup V(G_0)
	\cup V(G_1) \rangle$ is colored with just one color. Thus, the remaining
	$k-1$ colors can only be assigned to vertices in complete subgraphs of
	$\mathcal K$. Since each such subgraph has $2m+k+1$ vertices and by Lemma
	\ref{subgraohW.col} it follows that all vertices within such a complete
	graph can only be colored with one color. Since there are only $k-2$ such
	complete graphs at most $k-2$ of the remaining $k-1$ colors can be used.
	Thus one might get a path-$(k-1)$-coloring, but no path-$k$-coloring.
	Hence, we can conclude $\mathcal C(v)\neq \mathcal C(w)$ for all vertices
	with coordinates $c^i(v) = (000)$ and $c^j(w) = (111)$.

%	Moreover, as argued before it holds $\mathcal C(x) = \mathcal C(y)$ for all vertices 
%	with coordinates $c^j(x)=c^i(y) \in {(000),(111)}$.
	The above statements and Lemma \ref{subgraohW.col} directly imply that
	the subgraph $G(m)$ and in particular all $Q^i\in \mathcal Q$ must be 
	path-$l$-colored using at	least two colors, i.e., $2\leq l \leq k$. 

	We continue to show that for any vertex $w \in V(\mathbb K^i)$ of each
	complete graph $\mathbb K^i\in \mathcal K$ and for all vertices $v\in
	V(Q^j)$ of each $Q^j\in \mathcal Q$ holds $\mathcal C(w)\neq\mathcal
	C(v)$.
	Assume first, there is a vertex $v\in V(Q^j)$ with coordinates $(000)$ or
	$(111)$ such that $\mathcal C(w) = \mathcal C(v)$ for some vertex $ w\in
	V(\mathbb K^i)$. Without loss of generality let $v$ have coordinates
	$(111)$. Let $u\in V(Q^j)$ be the vertex with coordinates $(000)$.
	Clearly, since $ \mathbb K^i$ is a complete subgraph of size $2m+k+1$ all
	vertices within $ \mathbb K^i$ must obtain the color $\mathcal C(v)$, in
	particular the vertex $x \in V(\mathbb K^i)$ that is connected to vertex
	$u$ via the edge $(u,x)\in F'$. By Lemma \ref{subgraohW.col} and \ref{lem:p24kQ3} 
	and since
	$\mathcal C(u)\neq \mathcal C(v)$ (as shown before), the subgraph $Q^j$
	can only be path-$2$-colored, path-$4$-colored or path-$8$-colored . If
	$Q^j$ is path-$2$-colored then Lemma \ref{lem:p2kQ3-1} implies that there
	is is a vertex $z\in N(u)$ with $\mathcal C(z) = \mathcal
	C(v)=\mathcal{C}(x)$ and hence the well-colored path $z-u-x$, a
	contradiction. If $Q^j$ is path-$4$-colored then Lemma \ref{lem:p2kQ3-1}
	implies that there is a vertex $z\in N(v)$ with
	$\mathcal{C}(z)=\mathcal{C}(v)=\mathcal{C}(x)$ and that there is a
	well-colored path from $z$ to $u$. Thus, there is a well-colored path
	from $z$ to $x$, using the edge $(u,x)$, a contradiction. If $Q^j$ is
	path-$8$-colored then any path within $Q^j$ from $v$ over $u$ to $x$ is
	well-colored, while $\mathcal{C}(v)=C(x)$, a contradiction. Hence, for
	all vertices $w \in V(\mathbb K^i)$ of each $\mathbb K^i\in \mathcal K$
	and for all vertices $v\in V(Q^j)$ with coordinates $(000)$ or $(111)$ of
	each $Q^j\in \mathcal Q$ holds $\mathcal C(w)\neq \mathcal C(v)$ . Now
	let $v\in V(Q^j)$ be an arbitrary vertex with coordinates different from
	$(000)$ or $(111)$ and assume that $\mathcal C(w)=\mathcal C(v)$ for for
	some vertex $w\in V(\mathbb K^i)$. Let $x$ and $y$ denote the vertices of
	$Q^j$ with coordinates $(000)$ and $(111)$, respectively. As argued
	before, each vertex within $ \mathbb K^i$ must obtain the color $\mathcal
	C(v)$, in particular the vertices $z, z' \in V(\mathbb K^i)$ that are
	connected to $x$, resp., $y$ via the edges $(z,x), (z',y)\in F'$.
 % Note, vertex $v$ is not connected to any vertex in any $\mathbb K^i$.
	Since $\mathcal C(x)\neq \mathcal C(y)$, $\mathcal C(x)\neq \mathcal C(w)
	= \mathcal C(v)$ and $\mathcal C(y)\neq \mathcal C(w) = \mathcal C(v)$
	this vertex $v$ must be colored with a third color $\mathcal C(w)$
	different from $\mathcal C(x)$ and $\mathcal C(y)$. However, vertex $v$
	has either distance one to vertex $x$ or to vertex $y$. Thus, there is
	always a well-colored path $v-x-z$ or $v-y-z'$, a contradiction.

	Finally, we show that for complete graphs $\mathbb K^i,\mathbb K^j \in
	\mathcal K$, $i\neq j$ holds $\mathcal C(v) \neq \mathcal C(w)$ for any
	vertex $v\in V(\mathbb K^i)$ and $w\in V(\mathbb K^j)$. Note that any
	path-$k$-coloring within each $\mathbb K^i$ must be constant since each
	$\mathbb K^i\in \mathcal K$ is of size $2m+k+1$. 
	If $\mathcal K = \{\mathbb  K^1\}$, i.e. if $k=3$, then there is nothing to show. 
	In particular, as argued before all vertices within the subgraph $\mathbb  K^1$
	must obtain a third color different from colors used in $G(m)$ and therefore,  
	by Lemma \ref{subgraohW.col}, 
	the subgraph $G(m)\subseteq G(m,3)$ must be path-$2$-colored whenever 
	$G(m,3)$ is path-$3$-colored. 
	Assume $|\mathcal K|\geq2$, i.e., $k>3$, and $\mathcal C(v) = \mathcal
	C(w)$ for $v\in V(\mathbb K^i)$, $w\in V(\mathbb K^j)$, $i\neq j$. Let
	$u\in V(Q^1)$ be the vertex with coordinates $(000)$. By construction, this
	vertex is connected to respective vertices $z\in V(\mathbb K^i)$ and
	$z'\in V(\mathbb K^j)$ via edges $(u,z),(u,z')\in F'$. 
	As shown before, $\mathcal C(u)\neq  \mathcal C(z)$
	and $\mathcal C(u)\neq  \mathcal C(z')$
   and hence, there is a well-colored path $z - u
	-z'$ with $\mathcal C(z) = \mathcal C(z')$, a contradiction. 

	To summarize, we have shown that for any path-$k$-coloring of the
	$k$-extended joint graph $G(m,k)$ each of the $k-2$ complete graphs in
	$\mathcal K$ must obtain different colors, while the coloring within each
	such complete graph is constant. Thus, $k-2$ of the $k$ colors are used.
	Moreover, none of the used $k-2$ colors can be reused in any $Q^j\in
	\mathcal Q$ and thus neither in $G(m)\subseteq G(m,k)$. Moreover, we
	showed that $\mathcal C(v)\neq \mathcal C(w)$ for all vertices with
	coordinates $c^j(v)= (000)$ and $c^j(v)=(111)$ and therefore the two
	unused colors must occur occur in all $Q^j\in \mathcal Q$ and hence in
	$G(m)\subseteq G(m,k)$. The above statements and Lemma
	\ref{subgraohW.col} directly imply that if  $G(m,k)$ 	is path-$k$-colored then
	the subgraph $G(m)$ must be	path-$2$-colored. 
\end{proof}

\subsection{Computational Complexity}

Now, we are able to prove the NP-completeness for deciding whether a graph
has a path-$k$-coloring. For this the next well-known Problem
\textsc{Monotone 1-in-3 SAT} and theorem will be crucial.

\begin{probl}\textsc{Monotone 1-in-3 SAT}\\
\begin{tabular}{ll}
	$\ \ \ $  \emph{Input:}&Given a set $U$ of Boolean variables and a set of clauses
	$\psi = \{L_1, \dots, L_m\}$ over $U$ 
	such that for all\\ 
	&$i=1, \dots, m$ holds: 
	$|L_i|=3$ and $L_i$ contains no negated variables.
	 \\
	$\ \ \ $\emph{Question:}&Is there a truth assignment to $\psi$ such that each $L_i$ contains
	exactly one true variable?
\end{tabular}
\end{probl}

\begin{thm}[\cite{Sch78}]
\textsc{Monotone 1-in-3 SAT} is NP-complete.
\end{thm}

\begin{probl}\textsc{Path-$k$-Coloring (P-$k$-Col)}\\
\begin{tabular}{ll}
	$\ \ \ $\emph{Input:}&Given an arbitrary connected graph $G = (V,E)$ 
				and an integer $K$ with $2 \leq K\leq |V|-1$. \\ %$|V|-1\geq K\geq 2$. \\
	 $\ \ \ $\emph{Question:}&	Is there a nontrivial path-$k$-coloring for $G$, $k\leq K$. % with $2 \leq k \leq K$? 
\end{tabular}
\end{probl}
%\ \\

We shortly summarize the main steps for proving the NP-completeness of
\textsc{P-$k$-Col}. After verifying that \textsc{P-$k$-Col} $\in$ NP, we
will show by reduction from \textsc{Monotone 1-in-3 SAT} that
\textsc{P-$k$-Col} is NP-hard: For a given instance $\psi = (L_1, \dots,
L_m)$ of \textsc{Monotone 1-in-3 SAT} we identify each clause $L_i$ with a
hypercube $Q^i\in \mathcal Q$. Moreover, each variable $b\in L_i$ is
identified with a unique vertex in $Q^i$. Finally, for an arbitrary integer
$k\geq 2$ we construct the $k$-extended joint graph $G(m, k)$ as in
Definition \ref{def:extGm} and show that $\psi$ has a truth assignment if
and only if $G(m, k)$ has a nontrivial path-$k$-coloring. Note, since $G(m,
2) = G(m)$ one can directly conclude that the problem of determining
whether a graph has a nontrivial path-$2$-coloring is NP-complete.

\begin{thm}
	\textsc{P-$k$-Col} is NP-complete.
\end{thm}
\begin{proof}
	First we show that \textsc{P-$k$-Col} $\in$ NP. Let $\mathcal{C}$ be a
	given $k$-coloring on $G=(V,E)$. We must show that one can verify in
	polynomial time if $\mathcal C$ is a path-$k$-coloring. For this, we
	first remove all edges $(u,v)\in E$ with $\mathcal{C}(u) =
	\mathcal{C}(v)$ and obtain a new graph $G'=(V,E')$, $E'\subseteq E$.
	Thus, for all edges $(u,v)\in E'$ holds $\mathcal{C}(u) \neq
	\mathcal{C}(v)$. Therefore, in each connected component of $G'$ all
	vertices are connected via a well-colored path. Hence, one has to verify
	that in none of the connected components of $G'$ some color occurs twice
	which can be done in polynomial time in the number of edges and vertices.

  We will show by reduction from \textsc{Monotone 1-in-3 SAT} that
  \textsc{P-$k$-Col} is NP-hard. %, even for $k=2$. 
	Let $\psi = (L_1, \dots, L_m)$ be an arbitrary instance of
	\textsc{Monotone 1-in-3 SAT}. Each clause $L_i$ will be identified with a
	hypercube $Q_3$ denoted by $Q^i$. Let $\mathcal{Q} = \{Q^1,\dots Q^m\}$
	be the set of these hypercubes. Each of the variables $b_1,b_2,b_3\in
	L_i$ is identified with one vertex $v_1,v_2,v_3 \in V(Q^i)$ that have
	coordinates $(001)$, $(010)$ and $(100)$, respectively. Different
	variables $b_r,b_s \in L_i$ are identified with different vertices. We
	can now create the joint graph $G(m)$ based on the set $\mathcal{Q}$ as
	in Definition \ref{def:Gm}, whereby we add an edge $(v,w) \in E'$ between
	two different hypercubes $Q^i$ and $Q^j$ if there are two variables $b\in
	L_i$ and $b'\in L_j$ with $b=b'$. For an arbitrary integer $k \geq 2$ we
	extend the graph $G(m)$ to the graph $G(m,k)$ as in Definition
	\ref{def:extGm}. Clearly, this reduction can be done in polynomial time
	in the number $m$ of clauses and the integer $k$. 

	We will show in the following that $G(m,k)$ has a path-$k$-coloring 
	if and only if $\psi$ has a truth assignment. 
  Let $\psi = (L_1, \dots, L_m)$ have a truth assignment. Thus, in each
  clause $L_i$ exactly one variable is true ($\boldsymbol{T}$) and two of
  the variables are false ($\boldsymbol{F}$). We will show that the
  corresponding joint graph $G(m)$ has a path-$2$-coloring
	$\mathcal{C}: V(G(m))  \rightarrow \{\mathbb{F},\mathbb{T}\}$,
	which than leads directly to a path-$k$-coloring  of $G(m,k)$.
  Thus, we first start to color the graph
  $G(m)\subseteq G(m,k)$. Let $W^j \subset V(Q^j)$ be the set of vertices
  contained in $G(m)$ with coordinates $c^j(v) \in \{(001),(010),(100)\}$.
  We assign to each vertex $v \in W^j $ the respective color $\mathbb{T}$
  or $\mathbb{F}$ according to the truth value of the unique Boolean
  variable contained in $L_j$ that is associated to vertex $v$. Hence,
  exactly two of the vertices contained in $W^j$ get color $\mathbb{F}$ and
  one gets color $\mathbb{T}$. Since all such vertices in $W^j$ are
  adjacent to the respective vertex $x \in V(Q^j)$ with coordinates $(000)$
  we must set $\mathcal{C}(x) = \mathbb{F}$. To obtain a proper
  path-$2$-coloring Lemma \ref{lem:p2kG=UnionQ3} implies that
  $\mathcal{C}(y) \neq \mathcal{C}(x)$ for the vertex $y$ with coordinates
  $(111)$, thus we set $\mathcal{C}(y) = \mathbb{T}$. The remaining
  vertices in $V(Q^j)$ can now easily be colored in a unique way regarding
  Lemma \ref{lem:G-k-color} in order to obtain a path-$2$-coloring of
  $\langle V(Q^j)\rangle$. All vertices in $G(m)$ are colored in this way
  w.r.t. to their corresponding induced subgraphs $\langle V(Q^j)\rangle$
  they are contained in. By construction, edges contained in $E_0$, resp.,
  $E_1$ connect only vertices with the same color $\mathbb{F}$, resp.,
  $\mathbb{T}$. The same holds for edges contained in $E'$, since $\psi$
  has a truth assignment and only those vertices are connected by edges
  $e\in E'$ if they represent the same variable in different clauses.
  Therefore, all well-colored paths of $G(m)$ are entirely included in the respective
  subgraphs $\langle V(Q^j)\rangle$, $j=1,\dots, m$ which are
  path-$2$-colored. Hence, the joint graph $G(m)$ has a path-$2$-coloring.
	In order to obtain a path-$k$-coloring of the the graph $G(m, k)$ we
	first observe that the vertices in $G_0$, must get color $\mathbb{F}$,
	since all vertices with coordinates $(000)$ induce together with the
	vertices in $G_0$ a complete graph $K_{k+1}$ and thus, only the trivial
	path-$1$-coloring can be used within this subgraph. Analogously, all 
	vertices in $G_1$ get color $\mathbb{T}$. By the same arguments,  
	each $\mathbb K^i \in \mathcal K$ can only be path-$1$-colored since 
	each $\mathbb K^i$ is of size $2m+k+1$ 	. Hence, we color each
	$\mathbb K^i \in \mathcal K$ with one of the remaining $k-2$ colors, 
	different graphs contained in $\mathcal K$ obtain different colors. 
	It is easy to see that the in this way new obtained well-colored paths
	can only be found in the induced subgraphs $\langle 
	\cup_{i=1}^{k-2} V(\mathbb K^i) \cup V(Q^l)\rangle$ for fixed $l= 1,\dots,m$. In
	particular, these new well-colored paths are either edges $(x,y) \in F'$
	or paths of the form $x-v-y$ with $x\in V(\mathbb K^i)$ for some 
	$\mathbb K^i \in \mathcal K$, 
	$v\in V(Q^l)$ and
	either $y\in V(Q^l)$ or $y\in V(\mathbb K^j)$, $j\neq i$, see also Figure
	\ref{fig:extjointgraph}. By construction, for all these cases holds
	$\mathcal C(x) \neq \mathcal C(y)$ and hence, we obtain a
	path-$k$-coloring of $G(m,k)$. 

%\\ 
%		the form $x'-v-w$ or $y'-v-w$ with with $x',y'\in \mathbb K^i_{k+1}$ for some $i$,
%	$v\in V(Q^1)$ and $w\in V(Q^1)$ or $w\in V(\mathbb K^j_{k+1})$, $j\neq
%	i$, see also Figure \ref{fig:extjointgraph}. By construction for all these
%	paths holds $\mathcal C(x') \neq \mathcal C(w)$ and $\mathcal C(y') \neq
%	\mathcal C(w)$ and hence, we obtain a path-$k$-coloring of $G(m.k)$. \\ 

	Conversely, suppose that $G(m,k)$	has a path-$k$-coloring $\mathcal C'$.
	We will show that the corresponding set
  of clauses $\psi = (L_1, \dots, L_m)$ has a truth assignment.	
	By Lemma \ref{lem:kcol-to-2col}, the restriction 	
	$\mathcal C = \mathcal C_{|V(G(m))}'$ is a path-$2$-coloring of $G(m)$.
	  Let $\mathcal{C}: V(G(m)) \rightarrow
  \{\mathbb{F},\mathbb{T}\}$ be such a path-$2$-coloring of the
  joint graph $G(m)\subseteq G(m,k)$.  Lemma
  \ref{lem:p2kG=UnionQ3} implies that $\mathcal{C}(v)\neq \mathcal{C}(w)$
  for all vertices $v,w \in V(G(m))$ with coordinates $c^j(v) = (000)$ and
  $c^i(w) = (111)$, while $\mathcal{C}(v) = \mathcal{C}(w)$ if $(v,w) \in
  E_0 \cup E_1$. Thus, w.l.o.g. we can assume that $\mathcal{C}(v) =
  \mathbb{F}$ and $\mathcal{C}(w)= \mathbb{T}$ for all vertices $v$ and $w$
  with coordinates $c^j(v) = (000)$ and $c^i(w) = (111)$, respectively. Let
  $v\in V(Q^j)$ be the vertex with coordinates
  $(000)$. Lemma \ref{lem:subgraph-k-coloring} and \ref{lem:p2kQ3-1} imply
  that there are two vertices in $N(v) \cap V(Q^j)$ with color $\mathbb{F}$
  and one with color $\mathbb{T}$ for all $Q^j\in \mathcal{Q}$ in $G(m)$.
  Note, the vertices in $N(v) \cap V(Q^j)$ have respective coordinates
  $(001)$, $(010)$, and $(100)$ and are by construction uniquely identified
  with the respective Boolean variables in $L_j$. Hence, we assign to each
  Boolean variable in each clause $L_j$ a value ``true'' ($\mathbb{T}$),
  resp., ``false''($\mathbb{F}$), depending on the color of the
  corresponding vertex in $V(Q^j)$. Therefore, two of the Boolean variables
  get the value $\mathbb{F}$ and one gets value $\mathbb{T}$ in each clause
  $L_j$. Finally, applying Lemma \ref{lem:p2kG=UnionQ3} again we can
  conclude that $\mathcal{C}(v) = \mathcal{C}(w)$ for all vertices
  $v,w$ with $(v,w) \in E'$. Those edges connect vertices whenever the
  corresponding Boolean variables in the different clauses are identical.
  Since those vertices share the same color, we can conclude that the
  assignment of the values $\mathbb{T}$, resp. $\mathbb{F}$, to the Boolean
  variables does not lead to any conflicts and therefore to a valid truth
  assignment for $\psi$. 
\end{proof}

We finish this contribution by stating several results concerning
the complexity of other problems that result from the previous theorem. 

\begin{probl}\textsc{Check S-composite}\\
\begin{tabular}{ll}
	$\ \ \ $\emph{Input:}&Given an arbitrary connected graph $G = (V,E)$. \\
	$\ \ \ $\emph{Question:}& Is $G$ S-composite? 
\end{tabular}
\end{probl}

\begin{thm}
\label{thm:NPc-Scomp}
	\textsc{Check S-composite} is NP-complete.
\end{thm}
\begin{proof}
To prove that \textsc{Check S-composite} $\in$ NP we must verify that a
proposed embedding of the given graph $G=(V,E)$ into a nontrivial
Cartesian product $G_1\Box G_2$ maps some edges of $G$ to copies of edges
of $G_1$ and some edges to copies of edges of $G_2$. This task can
obviously be done in polynomial time in the number of edges of $G$.

The NP-hardness follows directly by using the same arguments as in the
proof of Theorem $1.$ in \cite{Klavzar:02}. The authors showed (indirectly)
the NP-hardness of \textsc{Check S-composite} by constructing a polynomial
time reduction from \textsc{P-$k$-col} and by proving that for any
path-$k$-coloring ($2\leq k \leq |V|-1$) a nontrivial embedding of $G$
into the Cartesian product of complete graphs 
$K_K\Box K_t$ ($t\geq2$) with $k\leq K$ can be found and
vice versa. 
\end{proof}

There are other characterizations of S-composite graphs as they can also be
defined in terms of edge labelings and nontrivial embeddings into Hamming
graphs, that is, Cartesian products of complete graphs.

\begin{thm}[\cite{KP:05}]
	Let $G=(V,E)$ be a connected graph. 
	The following statements are equivalent:
	\begin{enumerate}
	\item $G$ is S-composite
	\item $G$ is $2$-labelable, i.e., $E$ can be labeled with two labels such that on any induced cycle of $G$
	on which both labels appear, the labels change at least three times while
	passing the cycle.
	\item  $G$ is a nontrivial subgraph of a Hamming graph with two
	factors.
	\end{enumerate} 
\end{thm}

Using the latter results and the constructions given in \cite{KP:05} 
we can formulate the following problems and theorem.

\begin{probl}\textsc{2-Labelable}\\
\begin{tabular}{ll}
	$\ \ \ $\emph{Input:}&Given an arbitrary connected graph $G = (V,E)$. \\
	$\ \ \ $\emph{Question:}& Is $G$ $2$-labelable? 
\end{tabular}
\end{probl}

\begin{probl}\textsc{2-Factor-Hamming Embedding}\\
\begin{tabular}{ll}
	$\ \ \ $\emph{Input:}&Given an arbitrary connected graph $G = (V,E)$. \\
	$\ \ \ $\emph{Question:}& Is $G$ a nontrivial subgraph of a Hamming graph 
									  with two factors? 
\end{tabular}
\end{probl}

\begin{thm}
	The Problems \textsc{2-Labelable} and \textsc{2-Factor-Hamming Embedding}
	are NP-hard.
\end{thm}

Finally, since \textsc{Check S-composite} is NP-complete
 its complementary decision problem \textsc{Check S-prime}
 is CoNP-complete.

\begin{probl}\textsc{Check S-prime}\\
\begin{tabular}{ll}
	$\ \ \ $\emph{Input:}&Given an arbitrary connected graph $G = (V,E)$. \\
	$\ \ \ $\emph{Question:}& Is $G$ S-prime? 
\end{tabular}
\end{probl}

\begin{thm}
	\textsc{Check S-prime} is CoNP-complete.
\end{thm}

\section*{Acknowledgment}
I thank Peter F. Stadler, Lydia Ostermeier, Daniel St\"{o}ckel and 
Jiong Guo for several useful remarks that helped to improve the presentation. This work
was supported in part by the \emph{Deutsche Forschungsgemeinschaft} (DFG)
Project STA850/11-1 within the EUROCORES Programme EuroGIGA (project
GReGAS) of the European Science Foundation.

%\section*{References}
\bibliographystyle{plain}
\bibliography{S_prime}

\begin{thebibliography}{10}

\bibitem{Bresar:03}
B.~Bre{\v{s}}ar.
\newblock On subgraphs of {C}artesian product graphs and {S}-primeness.
\newblock {\em Discrete Math.}, 282:43--52, 2004.

\bibitem{GJ79}
M.R. Garey and D.S. Johnson.
\newblock {\em Computers and intractability}, volume 174.
\newblock Freeman San Francisco, CA, 1979.

\bibitem{Hammack:2011a}
R.~Hammack, W.~Imrich, and S.~Klav{\v{z}}ar.
\newblock {\em Handbook of Product Graphs}.
\newblock Discrete Mathematics and its Applications. CRC Press, 2nd edition,
  2011.

\bibitem{Hellmuth:11}
M.~Hellmuth.
\newblock A local prime factor decomposition algorithm.
\newblock {\em Discrete Math.}, 311(12):944--965, 2011.

\bibitem{HL:16}
M.~Hellmuth and F.~Lehner.
\newblock Fast factorization of {C}artesian products of (directed) hypergraphs.
\newblock {\em J. Theor. Comp. Sci.}, 615:1--11, 2016.

\bibitem{HON-14}
M.~Hellmuth, L.~Ostermeier, and M.~Noll.
\newblock Strong products of hypergraphs: Unique prime factorization theorems
  and algorithms.
\newblock {\em Discr. Appl. Math.}, 171:60--71, 2014.

\bibitem{HOS12}
M.~Hellmuth, L.~Ostermeier, and P.F. Stadler.
\newblock Diagonalized {C}artesian products of {S}-prime graphs are {S}-prime.
\newblock {\em Discrete Math.}, 312(1):74 -- 80, 2012.
\newblock Algebraic Graph Theory — A Volume Dedicated to Gert Sabidussi on
  the Occasion of His 80th Birthday.

\bibitem{IMKL-00}
W.~Imrich and S~Klav{\v{z}}ar.
\newblock {\em Product graphs}.
\newblock Wiley-Interscience Series in Discrete Mathematics and Optimization.
  Wiley-Interscience, New York, 2000.

\bibitem{IMKLDO-08}
W.~Imrich, S.~Klav{\v{z}}ar, and F.~D. Rall.
\newblock {\em Topics in Graph Theory: Graphs and Their Cartesian Product}.
\newblock AK Peters, Ltd., Wellesley, MA, 2008.

\bibitem{IP:07}
W.~Imrich and I.~Peterin.
\newblock Recognizing {C}artesian products in linear time.
\newblock {\em Discrete Math.}, 307(3-5):472--483, 2007.

\bibitem{Klavzar:02}
S.~Klav{\v{z}}ar, A.~Lipovec, and M.~Petkov{\v{s}}ek.
\newblock On subgraphs of {C}artesian product graphs.
\newblock {\em Discrete Math.}, 244:223--230, 2002.

\bibitem{KP:05}
S.~Klav{\v{z}}ar and I.~Peterin.
\newblock Characterizing subgraphs of {H}amming graphs.
\newblock {\em J. Graph Theory}, 49(4):302--312, 2005.

\bibitem{LB:81}
R.~H. Lamprey and B.~H. Barnes.
\newblock A new concept of primeness in graphs.
\newblock {\em Networks}, 11:279--284, 1981.

\bibitem{LB:95}
R.~H. Lamprey and B.~H. Barnes.
\newblock A characterization of {C}artesian-quasiprime graphs.
\newblock {\em Congressus Numerantium}, 109:117--121, 1995.

\bibitem{Sab:75}
G.~Sabidussi.
\newblock Subdirect representations of graphs in infinite and finite sets.
\newblock {\em Colloq. Math. Soc. Janos Bolyai}, 10:1199--1226, 1975.

\bibitem{Sch78}
T.J. Schaefer.
\newblock The complexity of satisfiability problems.
\newblock In {\em Proceedings of the tenth annual ACM symposium on Theory of
  computing}, STOC '78, pages 216--226, New York, NY, USA, 1978. ACM.

\end{thebibliography}

% ------------------------------------------------------------------------

%\subsection*{Acknowledgment}
%Many thanks to our \TeX-pert for developing this class file.
% ------------------------------------------------------------------------
\end{document}